\documentclass[a4paper,10pt]{article}
\usepackage[english]{babel}
\usepackage[utf8]{inputenc}
\usepackage{amsfonts}
\usepackage{amsthm}
\usepackage{amsmath}
\usepackage{amssymb}
\usepackage{enumerate}
\usepackage{mathtools}
\usepackage{tikz}
\usepackage{graphicx}
\usepackage{tikz-cd}
\usepackage{tcolorbox}
\usepackage{authblk}
\usepackage{rotating}
\usepackage{mathtools}

\definecolor{myurlcolor}{rgb}{0.6,0,0}
\definecolor{mycitecolor}{rgb}{0,0,0.8}
\definecolor{myrefcolor}{rgb}{0,0,0.8}
\usepackage[bookmarks=false]{hyperref}
\hypersetup{colorlinks, linkcolor=myrefcolor, citecolor=mycitecolor, urlcolor=myurlcolor}

\usetikzlibrary{arrows,backgrounds,circuits,circuits.ee.IEC,shapes,fit,matrix,decorations.markings, positioning}
\tikzstyle{none}=[inner sep = 0]

\definecolor{lblue}{rgb}{0,250,255}
\tikzstyle{species}=[circle,fill=yellow,draw=black,scale=2.15, inner sep = 2pt]
\tikzstyle{reaction}=[rectangle,fill=lblue,draw=black,scale=2]
\tikzstyle{inarrow}=[->, >=stealth, shorten >=.03cm,line width=1.5]
\tikzstyle{empty}=[circle,fill=none, draw=none]
\tikzstyle{inputdot}=[circle,fill=purple,draw=purple, scale=.25]
\tikzstyle{inputarrow}=[->,draw=purple, shorten >=.05cm]
\tikzstyle{simple'}=[-,draw=purple,line width=1.000]
\tikzstyle{arrow}=[-,draw=black,postaction={decorate},decoration={markings,mark=at position .5 with {\arrow{>}}},line width=1.200]

\pgfdeclarelayer{edgelayer}
\pgfdeclarelayer{nodelayer}
\pgfsetlayers{edgelayer,nodelayer,main}

\tikzstyle{main node} =[circle,fill=white!20,draw,font=\sffamily\Large\bfseries]
\tikzstyle{terminal}=[circle,fill=white!20,draw,font=\sffamily\Large\bfseries,color=purple,fill=none]
\tikzstyle{none}=[inner sep=0pt]
\tikzstyle{dot}=[circle,fill=black,draw=black]
\tikzstyle{simple}=[-,draw=black,line width=1.500]

\usetikzlibrary{arrows}
\usetikzlibrary{decorations.markings}

\usepackage{color}

\newcommand{\tc}{\textcolor{blue}}
\newcommand{\btk}{\begin{tikzcd}}
\newcommand{\etk}{\end{tikzcd}}

\newcommand{\A}{\text{A}}
\newcommand{\B}{\text{B}}
\newcommand{\C}{\text{C}}
\newcommand{\ATP}{\text{ATP}}
\newcommand{\ADP}{\text{ADP}}
\newcommand{\AMP}{\text{AMP}}

\newcommand{\Ph}{\text{P}_{\text{i}}}
\newcommand{\PPh}{\text{PP}_{\text{i}}}
\newcommand{\Y}{\text{Y}}
\newcommand{\XY}{\text{XY}}
\newcommand{\X}{\text{X}}
\newcommand{\XPi}{\text{XP}_{\text{i}}}
\newcommand{\water}{\text{H}_2\text{O}}
\newcommand{\CoASH}{\text{CoA-SH}}
\newcommand{\aCoA}{\text{acetyl-CoA}}
\newcommand{\NAD}{\text{NAD}}
\newcommand{\FAD}{\text{FAD}}

\renewcommand{\O}{\text{O}}
\renewcommand{\H}{\text{H}}
\newcommand{\N}{\text{N}}

\newcommand{\acetate}{\C_2\H_3\O_2^-}
\newcommand{\fumarate}{\text{fumarate}}
\newcommand{\urea}{\text{urea}}
\newcommand{\aspartate}{\text{aspartate}}
\newcommand{\carbamoyl}{\text{carbamoyl phosphate}}

\newcommand{\myleftrightharpoons}{\ \xrightleftharpoons{\hspace{0.5cm}} \ }

\newcommand{\flylongleftrightharpoons}[1]{ \xrightleftharpoons[ \ #1_\leftarrow \ ]{#1_\rightarrow} }
\newcommand{\flyleftrightharpoons}[1]{ \xrightleftharpoons[ #1_\leftarrow ]{#1_\rightarrow} }

\newcommand{\updownharpoons}{\rotatebox[origin=c]{-90}{$\ \rightleftharpoons \ $}}

\begin{document}

\title{Biochemical Coupling Through \\ Emergent Conservation Laws}



\author{{\small JOHN C. BAEZ, BLAKE S. POLLARD, JONATHAN LORAND,\\ \vspace{-0.4cm}AND MARU SARAZOLA}}
\maketitle

\begin{abstract}
\noindent
Bazhin has analyzed ATP coupling in terms of quasiequilibrium states where fast reactions
have reached an approximate steady state while slow reactions have not yet reached
equilibrium.   After an expository introduction to the relevant aspects of reaction network
theory, we review his work and explain the role of ``emergent conservation laws'' in 
coupling.  These are quantities, left unchanged by fast reactions, whose conservation
forces exergonic processes such as ATP hydrolysis to drive desired endergonic processes. 
\end{abstract}

\section{Introduction}
\label{sec:introduction}

In the cell, chemical reactions are often ``coupled'' so that reactions that release energy drive
reactions that are biologically useful but involve an increase in energy.   But how, exactly,
does coupling work?

Much is known about this question, but the literature is also full of vague explanations
and oversimplifications.   Coupling cannot occur in equilibrium; it arises
in open systems, where the concentrations of certain chemicals are held out of equilibrium
due to flows in and out \cite{BP,EP,Pr}.    One might thus suspect that the simplest mathematical treatment of this phenomenon would involve non-equilibrium steady states of open systems.   However, Bazhin \cite{Bazhin} has shown that some crucial aspects of coupling  arise in an even simpler framework.   He considers ``quasi-equilibrium'' states, where fast reactions have come into equilibrium and slow ones are neglected.   He shows that
coupling occurs already in this simple approximation.

Our goal here is two-fold.  First,
we review Bazhin's work in a way that readers with no training in biology or chemistry can follow.  Second, we explain a fact that seems to have received little attention: in many cases, coupling relies on \emph{emergent  conservation laws}.

Conservation laws are important throughout science.   Besides those
that are built into the fabric of physics, such as conservation of energy and
momentum, there are also many ``emergent'' conservation laws that hold approximately
in certain circumstances.  Often these arise when processes that change a given quantity
happen very slowly.   For example, the most common isotope of uranium decays into lead
with a half-life of about 4 billion years---but for the purposes of chemical experiments in the laboratory, it is useful to treat the amount of uranium as a conserved quantity.

The emergent conservation laws involved in biochemical coupling are of a different
nature.  Instead of making the processes that violate these laws happen \emph{more slowly},
the cell uses enzymes to make other processes happen \emph{more quickly}.   At the
time scales relevant to cellular metabolism, the fast processes dominate, while
slowly changing quantities are effectively conserved.  By a suitable choice of these emergent
conserved quantities, the cell ensures that certain reactions that release energy can only
occur when other ``desired'' reactions occur.  To be sure, this is only \emph{approximately}
true, on sufficiently short time scales.    But this approximation is enlightening.

Following Bazhin, our main example involves coupling to ATP hydrolysis.  We consider the
following schema, which abstractly describes a whole family of reactions:
\begin{align}
\label{coupled1}
\X + \ATP  & \myleftrightharpoons \ADP + \XPi \\
\label{coupled2}
\XPi +\Y  & \myleftrightharpoons \XY + \Ph \end{align}
Some concrete examples of this schema include:
\begin{itemize}
\item The synthesis of glutamine ($\XY$) from glutamate ($\X$) and ammonium ($\Y$).
This is part of the important glutamate-glutamine cycle in the central nervous system.
\item The synthesis of sucrose ($\XY$) from glucose ($\X$) and fructose ($\Y$).  This is one
of many processes whereby plants synthesize more complex sugars and starches from simpler
building-blocks.
\end{itemize}
In these and other examples, reactions (\ref{coupled1}) and (\ref{coupled2}), taken together, have the effect of synthesizing a larger molecule $\XY$ out of two parts $\X$ and $\Y$, while $\ATP$ is broken down to $\ADP$ and $\Ph$.    Thus, they have the same net effect as this other pair of reactions:
\begin{align}
\label{uncoupled1}
 \X + \Y & \myleftrightharpoons  \XY
 \\
 \label{uncoupled2}
\ATP & \myleftrightharpoons\ \ADP + \Ph
\end{align}
where the desired reaction (\ref{uncoupled1})  involves an energy increase, while (\ref{uncoupled2}), a deliberately simplified version of ATP hydrolysis, releases energy.  But in reactions (\ref{coupled1}) and (\ref{coupled2}), these reactions are ``coupled'' so that $\ATP$ can only break down to $\ADP + \Ph$ if $\X + \Y$ also turns into $\XY$.

As we shall see, this coupling crucially relies on a conserved quantity: the total number of
$\Y$ molecules plus the total number of $\Ph$'s is left unchanged by both reactions above.  This fact is not a fundamental law of physics, nor even an approximate law considered fundamental in chemistry, such as conservation of phosphorus. It is an \emph{emergent} conservation law that holds approximately in special situations.   Its approximate validity relies on the fact that the cell has enzymes that make the two reactions in the above schema occur more rapidly than reactions that violate this law, such as these:

In what follows, Section \ref{sec:background} provides the background in chemistry needed
to follow the paper.  Section \ref{sec:coupling} poses the question ``what is coupling?''
Section \ref{sec:interactions} introduces the new reactions required for coupling ATP hydrolysis to the synthesis of $\XY$ from components $\X$ and $\Y$, and it explains
why these reactions are not yet enough for coupling.  Section \ref{sec:quasiequilibria}
shows that coupling occurs in a ``quasiequilibrium'' state where these new reactions,
assumed much faster than the rest, have reached equilibrium, while the rest are neglected.   Section \ref{sec:conservation} explains the importance of emergent conservation laws.   Finally, Section \ref{sec:another} shows that the same pattern is at work in a very different example of coupling: the 
urea cycle.

\section{Some background}
\label{sec:background}

In what follows, we will be working with reaction networks.  A reaction network consists of a set of {\it reactions}, for example
\[
\X+\Y\longrightarrow \XY.
\]
Here X, Y and XY are the {\it species} involved, and we interpret this reaction as species X and Y combining to form species XY.  We call X and Y the {\it reactants} and XY the {\it product}.  Additive combinations of species, such as $\X + \Y$, are called \emph{complexes}.

The \emph{law of mass action} states that the rate at which a reaction occurs is proportional to the product of the concentrations of the reactants. The proportionality constant is called the {\it rate constant}; it is a positive real number associated to a reaction that depends on chemical properties of the reaction along with the temperature, the pH of the solution, the nature of any catalysts that may be present, and so on.   Every reaction has a \emph{reverse} reaction; that is, if X and Y combine to form XY, then XY can also split into X and Y.   The reverse reaction has its own rate constant.

We summarize this information by writing
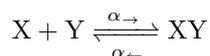
\[
 \X + \Y \flylongleftrightharpoons{\alpha}  \XY
\]
where $\alpha_{\to}$ is the rate constant for X and Y to combine and form XY, while
$\alpha_\leftarrow$ is the rate constant for the reverse reaction.

As time passes and reactions occur, the concentration of each species will likely change. We can record this information in a collection of functions $[\X] \colon \mathbb{R} \to [0,\infty)$, one for
each species $\X$, where $[\X](t)$ gives the concentration of the species $\X$ at time $t$. This naturally leads one to consider the {\it rate equation} of a given reaction, which specifies the time evolution of these concentrations.  The rate equation can be read off from the reaction network, and in the above example it is:
\begin{align*}
\dot{[\X]} & = -\alpha_\to [\X][\Y]+\alpha_\leftarrow [\XY]\\
\dot{[\Y]} & = -\alpha_\to [\X][\Y]+\alpha_\leftarrow [\XY]\\
\dot{[\XY]} & = \alpha_\to [\X][\Y]-\alpha_\leftarrow [\XY]
\end{align*}
Here $\alpha_\to [\X] [\Y]$ is the rate at which the forward reaction is occurring.
Similarly, $\alpha_\leftarrow [\XY]$ is the rate at
which the reverse reaction is occurring.

We say that a system is in {\it detailed balanced equilibrium}, or simply {\it equilibrium},
when every reaction occurs at the same rate as its reverse reaction.  This implies that the concentration of each species is constant in time.   In our example, the condition for equilibrium
is
\[
\frac{\alpha_\to}{\alpha_\leftarrow}=\frac{[\XY]}{[\X][\Y]}
\]
and the rate equation then implies that $\dot{[\X]} =  \dot{[\Y]} = \dot{[\XY]} = 0$.

The laws of thermodynamics determine the ratio of the forward and reverse rate constants.  For any reaction whatsoever, this ratio is
\begin{equation}
\label{gibbs}
     \frac{\alpha_\to}{\alpha_\leftarrow} = e^{-\Delta {G^\circ}/RT}
 \end{equation}
where $T$ is the temperature, $R$ is the ideal gas constant, and $\Delta {G^\circ}$ is the  free energy change under standard conditions.

Note that if $\Delta {G^\circ} < 0$, then $\alpha_\to > \alpha_\leftarrow$: the rate constant of the forward reaction is larger than the rate constant of the reverse reaction.  In this case one may loosely say that the forward reaction ``wants'' to happen ``spontaneously".  Such a reaction is called \emph{exergonic}.   If on the other hand $\Delta {G^\circ} > 0$, then the forward reaction is ``non-spontaneous" and it is called \emph{endergonic}.

The most important thing for us is that  $\Delta {G^\circ}$ takes a very simple form.    Each species has a free energy.   The free energy of a complex $\A_1 + \cdots + \A_m$ is the sum of the free energies of the species $\A_i$.   Given a reaction
\[  \A_1 + \cdots + \A_m \longrightarrow \B_1 + \cdots + \B_n, \]
the free energy change $\Delta {G^\circ}$ for this reaction is the free energy of
$\B_1 + \cdots + \B_n$ minus the free energy of  $\A_1 + \cdots + \A_m$.

As a consequence, $\Delta{G^\circ}$ is additive with respect to combining multiple reactions in either series or parallel.  In particular, then, the law (\ref{gibbs}) imposes relations between ratios of rate constants: for example, if we have the following more complicated set of reactions
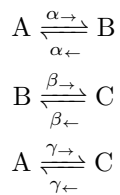
\begin{align*} \A & \flyleftrightharpoons{\alpha} \B \\ 
 \B & \flyleftrightharpoons{\beta} \C \\
 \A & \flyleftrightharpoons{\gamma} \C 
\end{align*}
then we must have
\[    \frac{\gamma_\to}{\gamma_\leftarrow} = \frac{\alpha_\to}{\alpha_\leftarrow}
\frac{\beta_\to}{\beta_\leftarrow} .  \]
So, not only are the rate constant ratios of reactions determined by differences in free energy, but also nontrivial relations between these ratios can arise, depending on the structure of the system of reactions in question.

\section{What is coupling?}
\label{sec:coupling}

Suppose that we are in a setting in which the reaction
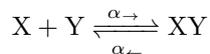
\[
 \X + \Y \flylongleftrightharpoons{\alpha}  \XY
\]
takes place. Let's also assume we are interested in the production of species XY from species X and Y, but that in our system, the reverse reaction is favored to happen.  This means that
\[  \alpha_\leftarrow \gg \alpha_\to \]
and so in equilibrium, the concentrations of the species will satisfy
\[  \frac{[\XY]}{[\X][\Y]}\ll 1 \]
which we assume undesirable.   How can we influence this ratio to get a more desired outcome?

This is where the notion of coupling comes into play. Informally, we think of the coupling of two reactions as a process in which an endergonic reaction---one which does not ``want" to happen---is combined with an exergonic reaction---one that \emph{does} ``want" to happen---in a way that improves the products-to-reactants concentrations ratio of the first reaction.

An important example of coupling, and one we will focus, on involves ATP hydrolysis
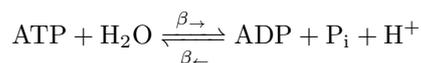
\[
\ATP + \water  \flylongleftrightharpoons{\beta} \ADP + \Ph + \H^+
\]
where ATP (adenosine triphosphate) reacts with a water molecule. Typically, this reaction results in ADP (adenosine diphosphate), a phosphate ion $\Pi$  and a hydrogen ion $\H^+$.  To simplify calculations, we will replace the above equation with
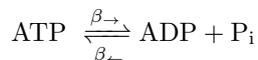
\[
\ATP  \ \flyleftrightharpoons{\beta} \ADP + \Ph
\]
since suppressing the bookkeeping of hydrogen and oxygen atoms in this manner will not affect our main points.

One reason ATP hydrolysis is good for coupling is that this reaction is strongly exergonic:
\[    \beta_\to \gg \beta_\leftarrow  \]
and in fact so much that
\begin{equation}
\label{big_ratio}
 \frac{\beta_\to}{\beta_\leftarrow} \gg \frac{\alpha_\leftarrow}{\alpha_\to}
\end{equation}
Yet this fact alone is insufficient to explain coupling.
To see why, suppose our system consists merely of the two reactions
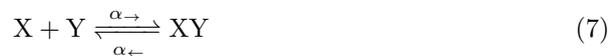
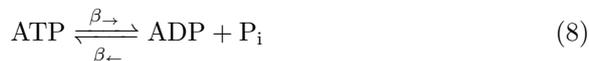
\begin{align}
\X + \Y   & \flylongleftrightharpoons{\alpha} \XY \label{alpha}\\[0.7em]
\ATP & \flylongleftrightharpoons{\beta} \ADP + \Ph \label{beta}
\end{align}
happening in parallel. We can study the concentrations in equilibrium to see that one reaction
has no influence on the other.   Indeed, the rate equation for this reaction network is
\begin{align*}
\dot{[\X]} & = -\alpha_\to [\X][\Y]+\alpha_\leftarrow [\XY]\\
\dot{[\Y]} & = -\alpha_\to [\X][\Y]+\alpha_\leftarrow [\XY]\\
\dot{[\XY]} & = \alpha_\to [\X][\Y]-\alpha_\leftarrow [\XY]\\
\dot{[\ATP]} & = -\beta_\to [\ATP]+\beta_\leftarrow [\ADP][\Ph]\\
\dot{[\ADP]} & = \beta_\to [\ATP]-\beta_\leftarrow [\ADP][\Ph]\\
\dot{[\Ph]} & = \beta_\to [\ATP]-\beta_\leftarrow [\ADP][\Ph]
\end{align*}
When concentrations are constant, these are equivalent to the relations
\[
\frac{[\XY]}{[\X][\Y]} = \frac{\alpha_\to}{\alpha_\leftarrow} \ \ \text{ and } \ \ \frac{[\ADP][\Ph]}{[\ATP]} = \frac{\beta_\to}{\beta_\leftarrow} \]
We thus see that ATP hydrolysis is in no way affecting the ratio of [XY] to [X][Y].   Intuitively, there
is no coupling because the two reactions proceed independently.  This ``independence'' is clearly visible if we draw the reaction network as a so-called Petri net \cite{BP}:
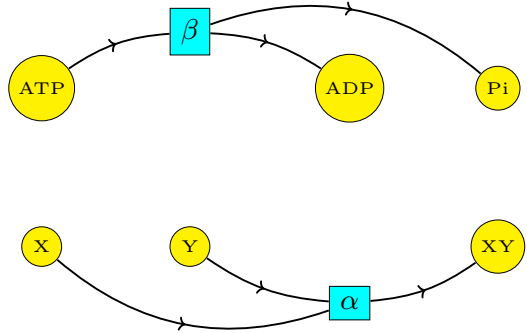
\[
\scalebox{0.6}{
\begin{tikzpicture}
	\begin{pgfonlayer}{nodelayer}
		\node [style=reaction] (0) at (1.75, -3) {\small{$\alpha$}};
		\node [style=species] (3) at (-5, 1.75) {\tiny{ATP}};
		\node [style=reaction] (5) at (-1.75, 3) {\small{$\beta$}};
		\node [style=species] (6) at (1.75, 1.75) {\tiny{ADP}};
		\node [style=species] (7) at (-1.75, -1.75) {\tiny{Y}};
		\node [style=species] (8) at (-5, -1.75) {\tiny{X}};
		\node [style=species] (9) at (5, -1.75) {\tiny{XY}};
		\node [style=species] (10) at (5,1.75) {\tiny{Pi}};
	\end{pgfonlayer}
	\begin{pgfonlayer}{edgelayer}
		\draw [style=arrow, bend right, looseness=1.00] (8) to (0);
		\draw [style=arrow, bend right=15, looseness=1.00] (7) to (0);
		\draw [style=arrow, bend left=15, looseness=1.00] (3) to (5);
		\draw [style=arrow, bend right=15, looseness=1.00] (0) to (9);
		\draw [style=arrow, bend left=15, looseness=1.00] (5) to (6);
		\draw [style=arrow, bend left, looseness=1.00] (5) to (10);
	\end{pgfonlayer}
\end{tikzpicture}
}
\]
So what really happens when we are in the presence of coupling?

\section{Interactions}
\label{sec:interactions}
For coupling to occur, the reactant species in both reactions must interact in some
way.  Indeed, in real-world examples that follow the above schema and involve coupling,
it is observed that, aside from the reactions represented in equations (\ref{alpha}) and
(\ref{beta}), two other reactions (and their reverses) take place:
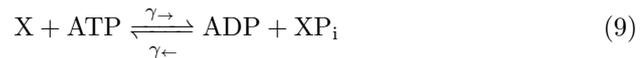
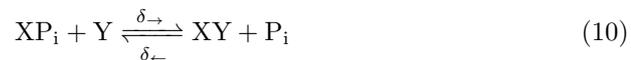
\begin{align}
\X + \ATP  & \flylongleftrightharpoons{\gamma} \ADP + \XPi \label{gamma}\\[0.7em]
\XPi +\Y  & \flylongleftrightharpoons{\delta} \XY + \Ph \label{delta} \end{align}
We can picture all four reactions (\ref{alpha})--(\ref{delta}) in a single Petri net as follows:
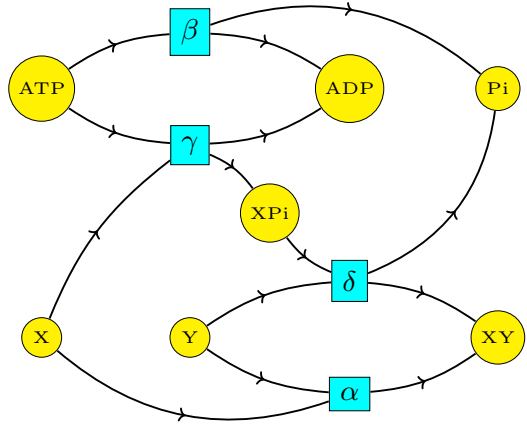
\[
\scalebox{0.6}{
\begin{tikzpicture}
	\begin{pgfonlayer}{nodelayer}
		\node [style=reaction] (0) at (1.75, -4) {\small{$\alpha$}};
		\node [style=reaction] (1) at (-1.75, 1.5) {\small{$\gamma$}};
		\node [style=species] (2) at (0, -0) {\tiny{XPi}};
		\node [style=species] (3) at (-5, 2.75) {\tiny{ATP}};
		\node [style=reaction] (4) at (1.75, -1.5) {\small{$\delta$}};
		\node [style=reaction] (5) at (-1.75, 4) {\small{$\beta$}};
		\node [style=species] (6) at (1.75, 2.75) {\tiny{ADP}};
		\node [style=species] (7) at (-1.75, -2.75) {\tiny{Y}};
		\node [style=species] (8) at (-5, -2.75) {\tiny{X}};
		\node [style=species] (9) at (5, -2.75) {\tiny{XY}};
		\node [style=species] (10) at (5, 2.75) {\tiny{Pi}};
	\end{pgfonlayer}
	\begin{pgfonlayer}{edgelayer}
		\draw [style=arrow, bend right=15, looseness=1.00] (3) to (1);
		\draw [style=arrow, bend left=15, looseness=1.00] (8) to (1);
		\draw [style=arrow, bend right, looseness=1.00] (8) to (0);
		\draw [style=arrow, bend right=15, looseness=1.00] (7) to (0);
		\draw [style=arrow, bend left=15, looseness=1.00] (3) to (5);
		\draw [style=arrow, bend left=15, looseness=1.00] (1) to (2);
		\draw [style=arrow, bend right=15, looseness=1.00] (2) to (4);
		\draw [style=arrow, bend right, looseness=1.00] (4) to (10);
		\draw [style=arrow, bend left=15, looseness=1.00] (4) to (9);
		\draw [style=arrow, bend right=15, looseness=1.00] (0) to (9);
		\draw [style=arrow, bend right=15, looseness=1.00] (1) to (6);
		\draw [style=arrow, bend left=15, looseness=1.00] (5) to (6);
		\draw [style=arrow, bend left, looseness=1.00] (5) to (10);
		\draw [style=arrow, bend left=15, looseness=1.00] (7) to (4);
	\end{pgfonlayer}
\end{tikzpicture}
}
\]

Taking into account this more complicated set of reactions, which are interacting with each other, is \emph{still not enough} to explain the phenomenon of coupling. To see this, consider the rate equation for the system comprised of all four reactions:
\begin{align*}
\dot{[\X]} & = -\alpha_\to [\X][\Y]+\alpha_\leftarrow [\XY]-\gamma_\to [\X][\ATP] + \gamma_\leftarrow [\ADP][\XPi]\\
\dot{[\Y]} & = -\alpha_\to [\X][\Y]+\alpha_\leftarrow [\XY]-\delta_\to [\XPi][\Y] +\delta_\leftarrow [\XY][\Ph]\\
\dot{[\XY]} & = \alpha_\to [\X][\Y]-\alpha_\leftarrow [\XY]+\delta_\to [\XPi][\Y] -\delta_\leftarrow [\XY][\Ph]\\
\dot{[\ATP]} & = -\beta_\to [\ATP]+\beta_\leftarrow [\ADP][\Ph] -\gamma_\to [\X][\ATP] + \gamma_\leftarrow [\ADP][\XPi]\\
\dot{[\ADP]} & = \beta_\to [\ATP]-\beta_\leftarrow [\ADP][\Ph] +\gamma_\to [\X][\ATP] - \gamma_\leftarrow [\ADP][\XPi]\\
\dot{[\Ph]} & = \beta_\to [\ATP]-\beta_\leftarrow [\ADP][\Ph]+\delta_\to [\XPi][\Y] -\delta_\leftarrow [\XY][\Ph]\\
\dot{[\XPi]} & = \gamma_\to [\X][\ATP] - \gamma_\leftarrow [\ADP][\XPi]-\delta_\to [\XPi][\Y] +\delta_\leftarrow [\XY][\Ph]
\end{align*}
Introducing the \textit{reaction velocities}
\begin{align*}
 J_\alpha &= \alpha_\to [\X][\Y] - \alpha_\leftarrow [\XY]  \\
 J_\beta  &= \beta_\to [\ATP] - \beta_\leftarrow [\ADP] [\Ph]  \\
 J_\gamma &= \gamma_\to [\ATP] [\X] - \gamma_\leftarrow [\ADP] [\XPi] \\
 J_\delta &= \delta_\to [\XPi] [\Y] - \delta_\leftarrow [\XY] [\Ph]
 \end{align*}
we can write the rate equation as
\begin{align*}
\dot{[\X]} & = -J_\alpha - J_\gamma  \\
\dot{[\Y]} & = -J_\alpha - J_\delta \\
\dot{[\XY]} & = J_\alpha + J_\delta \\
\dot{[\ATP]} & = -J_\beta - J_\gamma \\
\dot{[\ADP]} & = J_\beta + J_\gamma \\
\dot{[\Ph]} & = J_\beta + J_\delta \\
\dot{[\XPi]} & = J_\gamma -J_\delta
\end{align*}
In a steady state, all these time derivatives are zero, so
\[ J_\alpha = J_\beta = -J_\gamma = - J_\delta \]
Furthermore, in a detailed balanced equilibrium every reaction occurs at
the same rate as its reverse reaction, so all four reaction velocities vanish.
This implies the relations
\[
\frac{[\XY]}{[\X][\Y]} = \frac{\alpha_\to}{\alpha_\leftarrow} , \quad
\frac{[\ADP][\Ph]}{[\ATP]} = \frac{\beta_\to}{\beta_\leftarrow}  , \quad
\frac{[\ADP] [\XPi]}{[\ATP][X]}  = \frac{\gamma_\to}{\gamma_\leftarrow}, \quad
\frac{[\XY][\Ph]}{[\XPi][Y]} = \frac{\delta_\to}{\delta_\leftarrow} .
\]

Thus, even when the reactants interact to form the new species $\XPi$, {\it there can be no coupling if the whole system is in equilibrium}, since then the ratio $[\XY]/[\X][\Y]$ is forced to be $\alpha_\to/\alpha_\leftarrow$.    Coupling can only arise out of equilibrium.   But how, precisely, does coupling occur?

\section{Quasiequilibria}
\label{sec:quasiequilibria}

Coupling is achieved through the action of enzymes.   An enzyme can increase the rate constant of a reaction.  However, it cannot change the ratio of forward to reverse rate constants, since that is fixed by the difference of free energies as in Equation (\ref{gibbs}), and the presence of an enzyme does not change this.   Indeed, if an enzyme \emph{could} change this ratio, there would be no
need for coupling!   Increasing the ratio $\alpha_+/\alpha_-$ in the reaction
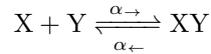
\[
 \X + \Y   \flylongleftrightharpoons{\alpha}  \XY
\]
would favor the formation of $\XY$, as desired.  But this option is not available.

Instead, the cell uses enyzmes to greatly increase the rate constants for (\ref{gamma}) and (\ref{delta}) while leaving those for (\ref{alpha}) and (\ref{beta}) essentially unchanged.    In this situation we can ignore reactions (\ref{alpha}) and (\ref{beta}) and still have a good approximate description of the dynamics, at least for sufficiently short time scales.   Thus, we
shall study \emph{quasiequilibria}, namely steady states of the rate equation for reactions
(\ref{gamma}) and (\ref{delta}) but not (\ref{alpha}) and (\ref{beta}).  In this approximation, the relevant Petri net becomes this:
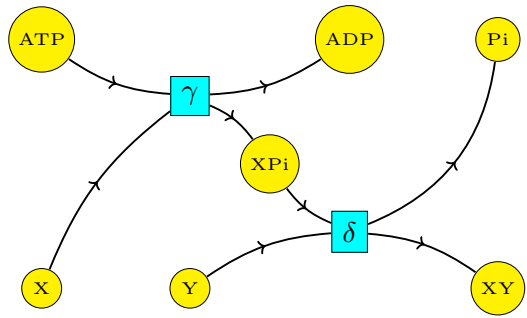
\[
\scalebox{0.6}{
\begin{tikzpicture}
	\begin{pgfonlayer}{nodelayer}

		\node [style=reaction] (1) at (-1.75, 1.5) {\small{$\gamma$}};
		\node [style=species] (2) at (0, -0) {\tiny{XPi}};
		\node [style=species] (3) at (-5, 2.75) {\tiny{ATP}};
		\node [style=reaction] (4) at (1.75, -1.5) {\small{$\delta$}};
		\node [style=species] (6) at (1.75, 2.75) {\tiny{ADP}};
		\node [style=species] (7) at (-1.75, -2.75) {\tiny{Y}};
		\node [style=species] (8) at (-5, -2.75) {\tiny{X}};
		\node [style=species] (9) at (5, -2.75) {\tiny{XY}};
		\node [style=species] (10) at (5, 2.75) {\tiny{Pi}};
	\end{pgfonlayer}
	\begin{pgfonlayer}{edgelayer}
		\draw [style=arrow, bend right=15, looseness=1.00] (3) to (1);
		\draw [style=arrow, bend left=15, looseness=1.00] (8) to (1);
		\draw [style=arrow, bend left=15, looseness=1.00] (1) to (2);
		\draw [style=arrow, bend right=15, looseness=1.00] (2) to (4);
		\draw [style=arrow, bend right, looseness=1.00] (4) to (10);
		\draw [style=arrow, bend left=15, looseness=1.00] (4) to (9);
		\draw [style=arrow, bend right=15, looseness=1.00] (1) to (6);
		\draw [style=arrow, bend left=15, looseness=1.00] (7) to (4);
	\end{pgfonlayer}
\end{tikzpicture}
}
\]
Now it is impossible for $\ATP$ to turn into $\ADP + \Ph$ without
$\X + \Y$ also turning into $\XY$.   As we shall see, this is the key to coupling.
In quasiequilibrium states, we shall find a nontrivial relation between the ratios $[\XY]/[\X][\Y]$ and $[\ATP]/[\ADP][\Ph]$.

Of course, this is just part of the full story.  Over longer time scales, reactions (\ref{alpha})
and (\ref{beta}) become important.   They would drive the system toward a true
equilibrium, and destroy coupling, if there were not an inflow of the reactants
$\ATP$, $\X$ and $\Y$ and an outflow of the products $\Ph$ and $\XY$.   To take
these inflows and outflows into account, we need the theory of `open' reaction networks
\cite{BP,EP,Pr}.

However, this is beyond our scope here.   We only consider reactions (\ref{gamma})
and (\ref{delta}), which give the following rate equation:
\begin{align*}
\dot{[\X]} & = -\gamma_\to [\X][\ATP] + \gamma_\leftarrow [\ADP][\XPi]\\
\dot{[\Y]} & = -\delta_\to [\XPi][\Y] +\delta_\leftarrow [\XY][\Ph]\\
\dot{[\XY]} & = \delta_\to [\XPi][\Y] -\delta_\leftarrow [\XY][\Ph]\\
\dot{[\ATP]} & =  -\gamma_\to [\X][\ATP] + \gamma_\leftarrow [\ADP][\XPi]\\
\dot{[\ADP]} & = \gamma_\to [\X][\ATP] - \gamma_\leftarrow [\ADP][\XPi]\\
\dot{[\Ph]} & = \delta_\to [\XPi][\Y] -\delta_\leftarrow [\XY][\Ph]\\
\dot{[\XPi]} & = \gamma_\to [\X][\ATP] - \gamma_\leftarrow [\ADP][\XPi]-\delta_\to [\XPi][\Y] +\delta_\leftarrow [\XY][\Ph].
\end{align*}
Quasiequilibria occur when all these time derivatives vanish.  This happens when
\begin{align*}
\gamma_\to [\X][\ATP] & = \gamma_\leftarrow [\ADP][\XPi]\\
\delta_\to [\XPi][\Y] & =\delta_\leftarrow [\XY][\Ph].
\end{align*}
This pair of equations is equivalent to
\[ \frac{\gamma_\to}{\gamma_\leftarrow}\frac{[\X][\ATP]}{[\ADP]}=[\XPi]
=\frac{\delta_\leftarrow}{\delta_\to}\frac{[\XY][\Ph]}{[\Y]} \]
and it implies
\[ \frac{[\XY]}{[\X][\Y]}
= \frac{\gamma_\to}{\gamma_\leftarrow}\frac{\delta_\to}{\delta_\leftarrow}
\frac{[\ATP]}{[\ADP][\Ph]}  \]
If we forget about the species  $\XPi$ (whose presence is crucial for the coupling to happen, but whose concentration we do not care about), the quasiequilibrium does not impose any conditions other than the above relation. We conclude that, under these circumstances and assuming we can increase the ratio of $[\ATP]$ to $[\ADP][\Ph]$, it is possible to increase the ratio of $[\XY]$ to $[\X][\Y]$.

We can say a bit more, since we can express the ratios of forward and reverse rate constants in terms of exponentials of free energy differences using equation (\ref{gibbs}).  Reactions (\ref{alpha}) and (\ref{beta}), taken together, convert $\X + \Y + \ATP$ to $\XY$ + ADP + Pi.   So do reactions (\ref{gamma}) and (\ref{delta}) taken together.  Thus, these two pairs of reactions involve the same
total change in free energy, so
\begin{equation}
\label{ratio_equation}
          \frac{\alpha_\to}{\alpha_\leftarrow}\frac{\beta_\to}{\beta_\leftarrow} =
 \frac{\gamma_\to}{\gamma_\leftarrow}\frac{\delta_\to}{\delta_\leftarrow} .
\end{equation}
Together with (\ref{big_ratio}), this implies
\[     \frac{\gamma_\to}{\gamma_\leftarrow}\frac{\delta_\to}{\delta_\leftarrow} \gg 1 \]
Thus,
 \[ \frac{[\XY]}{[\X][\Y]} \gg \frac{[\ATP]}{[\ADP][\Ph]}.  \]
This is why coupling to ATP hydrolysis is so good at driving the synthesis of $\XY$ from
$\X$ and $\Y$.   Ultimately, this inequality arises from the fact that the \emph{decrease} in free energy for the  reaction $\ATP \to \ADP + \Ph$ greatly exceeds the \emph{increase} in free energy
for the reaction $\X + \Y \to \XY$. But this fact can only be ``put to use'' in the presence
of coupling; that is, when we observe interactions as described in section \ref{sec:interactions} that are in a quasiequilibrium as explained in this present section.

\section{Emergent conservation laws}
\label{sec:conservation}

The rate equations for reactions (\ref{gamma}--\ref{delta}) have the following \emph{conserved quantities}; in other words, the following quantities are constant in time.
\begin{enumerate}
\item\label{law1} $[\X] + [\XPi] + [\XY]$, due to the conservation of X,
\item $[\Y] + [\XY]$, due to the conservation of Y,
\item $3[\ATP] +[\XPi] +[\Ph] +2[\ADP]$, due to the conservation of phosphorus,
\item\label{law4} $[\ATP] + [\ADP]$, due to the conservation of adenosine.
\end{enumerate}

In fact, these conserved quantities were already present in the  larger system involving reactions (\ref{alpha}-\ref{delta}). Moreover, these quantities, and their linear combinations, are the {\it only} conserved quantities for the larger system.

To see this, we use some standard ideas from reaction network theory \cite{F,HJ}.
Consider the 7-dimensional space with orthonormal basis  $\ATP, \ADP, \Ph, \XPi,$ $\X,$ $\Y,$ and $\XY$.  We  can think of complexes as vectors in this space.  An arbitrary choice of the concentrations of all species also defines a vector in this space.    Furthermore, any reaction
involving these species defined a vector in this space, namely the sum of the products minus
the sum of the reactants.  This is call the {\it reaction vector} of this reaction.   The four reactions (\ref{alpha}-\ref{delta}) give these reaction vectors:
\begin{align*}
 v_\alpha &= \XY - \X - \Y  \\
v_\beta &=  \Ph + \ADP - \ATP \\
v_\gamma &= \XPi + \ADP -  \ATP - \X \\
v_\delta &=  \XY + \Ph -  \XPi -  \Y
\end{align*}

Any change in concentrations caused by these reactions must lie in the {\it stoichiometric subspace}: that is, the space spanned by the reaction vectors.   Since these vectors obey one nontrivial relation, $v_\alpha + v_\beta = v_\gamma + v_\delta$, the stochiometric subspace is 3-dimensional.   Therefore, the space of conserved quantities must be 4-dimensional, since these specify the constraints on allowed changes in concentrations.

In contrast to this situation, if we consider only reactions (\ref{beta}) and (\ref{gamma}), the stoichiometric subspace is 2-dimensional, since $v_\beta$ and $v_\gamma$ are linearly independent.
Thus, the space of conserved quantities becomes 5-dimensional.   Indeed, we can find an additional conserved quantity:
\begin{enumerate}
\item[5.] $[\Y ] +[\Ph ]$
\end{enumerate}
that is linearly independent from the four conserved quantities we had before. It does not derive from the conservation of a particular molecular component.  In other words, conservation of this quantity is not a fundamental law of chemistry.
Instead, it is an {\it emergent} conservation law
that arises only in situations where the rate constants
of reactions (\ref{gamma}) and (\ref{delta}) are so much larger than those of (\ref{alpha})
and (\ref{beta}) that we can ignore the latter two reactions.

This emergent conservation law captures the phenomenon of coupling.   The only way for $\ATP$ to form $\ADP + \Ph$ without violating this law is for $\Y$ to be consumed in the same amount as $\Ph$ is created, thus forming the product $\XY$.

\section{Another example}
\label{sec:another}

ATP hydrolysis is a simple example of coupling through emergent conservation laws, but the phenomenon is more general.  A slightly more complicated example is the urea cycle.
The first metabolic cycle to be discovered, it is used by land-dwelling vertebrates
to convert ammonia, which is highly toxic, to urea for excretion.

The urea cycle consists of these reactions:
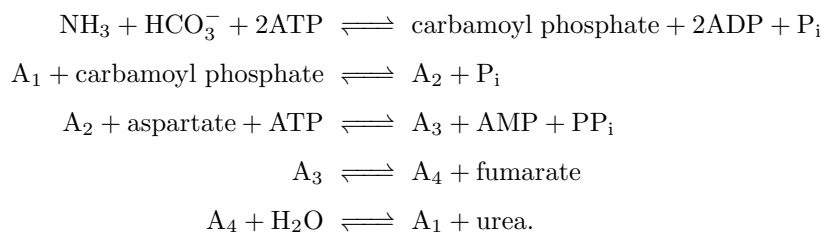
\begin{align*}
 \N\H_3 + \H\C\O_3^- + 2 \ATP &\myleftrightharpoons \carbamoyl + 2 \ADP + \Ph   \\
 \A_1 + \carbamoyl &\myleftrightharpoons \A_2 + \Ph \\
 \A_2 + \aspartate + \ATP &\myleftrightharpoons \A_3 + \AMP + \PPh \\
 \A_3  &\myleftrightharpoons \A_4 + \fumarate  \\
 \A_4 + \water &\myleftrightharpoons \A_1 + \urea.
\end{align*}
Ammonia ($\N\H_3$) and carbonate ($\H\C\O_3^-$) enter in the first reaction, along with $\ATP$.   The four remaining reactions form a cycle in which various species $\A_1, \dots ,\A_4$ ``cycle around,'' each transformed into the next\footnote{For the curious reader, these species are $\A_1=$ ornithine, $\A_2 =$ citrulline, $\A_3 =$ argininosuccinate,  and $\A_4 =$ arginine.}.   One atom of nitrogen from carbamoyl phosphate
and one from aspartate enter this cycle, and they are incorporated in urea, which
leaves the cycle along with fumarate.

All this is powered by two exergonic reactions: the hydrolysis of $\ATP$ to $\ADP$ and $\Ph$ and the hydrolysis of $\ATP$ to adenosine monophosphate ($\AMP$) and a compound with two phosphorus atoms, pyrophosphate ($\PPh$).   Thus, we are seeing a more elaborate example of an endergonic process coupled to $\ATP$ hydrolysis.  The most interesting new feature is the use of a cycle.

Since inflows and outflows are crucial to the purpose of the urea cycle, a full analysis requires treating this cycle as an open chemical reaction network \cite{BP,EP,Pr}.    However, we can gain some insight into coupling just by studying the emergent conservation laws present in this network, ignoring inflows and outflows.

There are a total of 16 species in the urea cycle.  There are 5 forward reactions, which
are easily seen to have linearly independent reaction vectors.  Thus, the stoichiometric
subspace has dimension 5.   There must therefore be 11 linearly independent conserved
quantities.

Some of these conserved quantities can be explained by fundamental laws of chemistry.
All the species involved are made of five different atoms: carbon, hydrogen, oxygen,
nitrogen and phosphorus.   The conserved quantity
\[  3[\ATP] + 2[\ADP] + [\AMP] + 2 [\PPh] + [\Ph] + [\carbamoyl]  \]
expresses conservation of phosphorus.  The conserved quantity
\[  [\N\H_3] + [\carbamoyl] +  [\aspartate] + 2[\urea] + 2[\A_1] + 3[\A_2] + 4[\A_3] + 4[\A_4] \]
expresses conservation of nitrogen.    Conservation of oxygen and carbon give still
more complicated conserved quantities.   Conservation of hydrogen and conservation
of charge are not really valid laws in this context, because all the reactions are occurring
in water, where it is easy for protons ($\H^+$) and electrons
to come and go.   So, four linearly independent ``fundamental'' conserved quantities are relevant to the urea cycle.

There must therefore be seven other linearly independent conserved quantities that
are ``emergent'': that is, not conserved in every possible chemical reaction, but
conserved by those in the urea cycle.  We can take these to be the following:

\begin{enumerate}
\item $[\ATP] + [\ADP] + [\AMP]$, due to conservation of adenosine by all reactions
in the urea cycle.
\item $[\water] + [\urea]$, since  the only reaction in the urea cycle involving
either $\water$ or $\urea$ has $\water$ as a reactant and $\urea$ as a product.
\item $[\aspartate] + [\PPh]$, since the only reaction involving
either $\aspartate$ or $\PPh$ has $\aspartate$ as a reactant and $\PPh$ as a product.
\item $2[\N\H_3] + [\ADP]$, since the only reaction involving either
$\N\H_3$ or $\ADP$ has $\N\H_3$ as a reactant and $2\ADP$ as a product
\item $2[\H\C\O_3^-] + [\ADP]$, since the only reaction involving either
$\H\C\O_3^-$ or $\ADP$ has $\H\C\O_3^-$ as a reactant and $2\ADP$ as a product.
\item $[\A_3] + [\fumarate] - [\PPh]$, since these species are involved only in the
third and fourth reactions of the urea cycle,
and this quantity is conserved in both those reactions.
\item $[\A_1] + [\A_2] + [\A_3] + [\A_4]$, since these species ``cycle around'' the
last four reactions, and they are not involved in the first.
\end{enumerate}

These emergent conservation laws prevent either form of ATP hydrolysis from occurring on its own: the reaction
\[     \ATP + \water \myleftrightharpoons \ADP + \Ph \]
violates conservation of quantities 2, 4 and 5, while
\[    \ATP + \water \myleftrightharpoons \AMP + \PPh \]
violates conservation of quantities 2, 3 and 6.   (In these reactions we are neglecting $\H^+$ ions, since as mentioned these are freely available in aqueous solution.)  Indeed, any linear combination of these two forms of ATP hydrolysis is prohibited.  Since this requires only
two emergent conservation laws, the presence of seven is a bit of a puzzle.
Conserved quantity 3 prevents the destruction of aspartate without the production
 of an equal amount of $\PPh$, conserved quantity 4 prevents the destruction of
 $\N\H_3$ without the production of an equal amount of $\ADP$, and so on.
But there seems to be more coupling than is strictly ``required''.   Of course, many factors
besides coupling are involved in an evolutionarily advantageous reaction network.

\subsubsection*{Acknowledgements}

We thank the organizers of Applied Category Theory 2018 for bringing us together to
work on this problem, and the Lorentz Center for hosting us.  We also thank our teammate
Fabrizio Genovese, and a helpful chemist on the \textsl{Azimuth} blog.

\noindent {\small \textsc{Department of Mathematics, University of California, Riverside CA, 92521, USA and Centre for Quantum Technologies, National University of Singapore, 117543, Singapore}}

{\it E-mail address:} \texttt{baez@math.ucr.edu}\\

\noindent {\small \textsc{Engineering Research Accelerator, Carnegie Mellon University, Pittsburgh PA, 15213, USA}}

{\it E-mail address:} \texttt{blake561@gmail.com}\\

\noindent {\small \textsc{Institute for Mathematics, University of Zurich, 8057 Zurich, Switzerland}}

{\it E-mail address:} \texttt{jonathanlorand@protonmail.com}\\

\noindent {\small \textsc{Department of Mathematics, Cornell University, Ithaca NY, 14853, USA}}

{\it E-mail address:} \texttt{mes462@cornell.edu}

\end{document}